# Novel phases in the Fe-Si-O system at terapascal pressures

Nan Huang[1], Renata M. Wentzcovitch[2,3,4,5,†], Zepeng Wu[1], Feng Zheng[6], Bingxin Wu[1], Yang Sun[1,‡], Shunqing Wu[1,*]

*[1]Department of Physics, OSED, Key Laboratory of Low Dimensional Condensed Matter Physics (Department of Education of Fujian Province), Xiamen University, Xiamen 361005, China*

*[2]Department of Applied Physics and Applied Mathematics, Columbia University, New York, NY 10027, USA*

*[3]Department of Earth and Environmental Sciences, Columbia University, New York, NY 10027, USA*

*[4]Lamont–Doherty Earth Observatory, Columbia University, Palisades, NY 10964, USA*

*[5]Data Science Institute, Columbia University, New York, NY 10027, USA*

*[6]School of Science, Jimei University, Xiamen, 361021, China*

## Abstract

The Fe-Si-O ternary system, central to modeling the interiors of terrestrial planets, remains poorly constrained at Terapascal (TPa) pressures characteristic of super-Earth mantles. Using a combination of crystal-structure prediction and *ab initio* calculations, we identify three ternary compounds stable near 1 TPa: $P3$ $FeSiO_4$, $P3$ $Fe_4Si_5O_{18}$, and $P\overline{3}$ $FeSi_2O_6$**.** The first two phases are thermodynamically stable at low temperatures, whereas $P\overline{3}$ $FeSi_2O_6$ becomes favored above ~2000 K. All three are metallic, paramagnetic, and adopt pseudo-binary arrangements derived from the $FeO_2$ and $SiO_2$ end-member structures. Their crystal structures emerge through substitutions of Fe for Si in $Fe_2P$-type $SiO_2$ or of Si for Fe in *Pnma*-type $FeO_2$, the stable elemental oxides at ~1 TPa. This structural continuity suggests that Fe preferentially substitutes for Si in the canonical Mg-silicates expected at TPa pressures. Notably, these new pseudo-binaries accommodate Fe in six- and nine-fold coordination, in contrast to the eight-fold cubic coordination found in FeO at similar pressures. The thermodynamic conditions under which these phases form from $FeO_2$ and $SiO_2$ mixtures are clarified through quasi-harmonic free-energy calculations. These phases imply a fundamentally different pattern of Fe incorporation into Mg-silicates at TPa pressures compared with that inferred for Earth's mantle, i.e., mainly $[Fe]_{Mg}$. Their stability may trigger silicate dissociation into oxides ($(Mg,Fe)_2(Si,Fe)O_4 \rightarrow 2(Mg,Fe)O + (Si,Fe)O_2$) at pressures below ~3 TPa, as expected in the Mg-Si-O system, with the extent of dissociation governed by iron content.

---

†Email: rmw2150@columbia.edu

‡Email: yangsun@xmu.edu.cn

*Email: wsq@xmu.edu.cn

## 1. Introduction

*Ab initio* materials simulations at extreme pressures and temperatures have reached a level of maturity that enables robust predictions of the physical properties of complex mineral assemblages and melts under conditions typical of planetary interiors, e.g., up to ~6,500 K and 3.6 Mbar in Earth and beyond (Umemoto et al., 2017). Earth remains the most comprehensively studied planet and the only one known to support life. Investigations of mantle-forming phases and their properties have provided critical insights into seismological observations, informing models of Earth's internal structure and dynamics.

The methodologies developed for these studies are now being extended to investigate other solar and extrasolar terrestrial materials, particularly those of super-Earth-type planets with masses of up to ~14 Earth masses ($M_\oplus$). These planets are being rapidly discovered and are of high interest due to their potential habitability. Comparative planetology, specifically the study of planetary interior structure, evolution, and dynamics, requires characterizing planet-forming phases under the extreme conditions expected in these bodies, which may reach temperatures of up to ~ 10,000 K and pressures of up to 3 TPa in terrestrial planets with masses of up to ~20 $M_\oplus$ (van den Berg et al., 2019; Fortney et al., 2007; Seager et al., 2007).

In the past two decades, the discovery of novel high-pressure phases has accelerated following the identification of the $MgSiO_3$ post-perovskite phase in 2004 (Tsuchiya et al., 2004; Murakami et al., 2004; Oganov et al., 2004), the highest-pressure Mg-silicate known in Earth's mantle. Coincidentally, the detection of the super-Earth 55 Cancri-e (McArthur et al., 2004) around the same time sparked interest in the Mg-Si-O system at high pressures, the dominant chemical system in rocky planetary mantles. All major computationally discovered phases in the Mg-Si-O system, up to 3 TPa (Wu et al., 2011; Wu et al., 2014; Umemoto et al., 2017; Umemoto et al., 2006; Umemoto et al., 2011; Niu et al., 2016; Tsuchiya et al., 2011; Liu et al., 2021), have also been predicted in low-pressure analogs, such as $NaMgF_3$ and $MgGeO_3$ (Umemoto et al., 2019; Umemoto et al., 2021) and subsequently confirmed experimentally (Grocholski et al., 2010; Dutta et al., 2019; Dutta et al., 2022). Notably, high-pressure phases often exhibit cation disorder (Umemoto et al., 2021; Dutta et al., 2022; Dutta et al., 2023).

Despite this progress, one of the major remaining challenges is incorporating iron, an abundant and geologically critical element, into these high-pressure systems. In Earth's lower mantle, iron exists primarily as $Fe^{2+}$ or $Fe^{3+}$, substituting for Mg and/or Si. In this work, we initiate the extension of the Mg-Si-O to include iron by exploring the stable Fe-Si-O ternary phases as possible end members at ~1 TPa. This investigation builds on extensive prior studies of the relevant binary systems, Fe-O (Zheng et al., 2022; Weerasinghe et al., 2015) and Si-O (Wu et al., 2011; Umemoto et al., 2019; Tsuchiya et al., 2011; Liu et al., 2021), at high pressures.

The following section describes the computational methods employed, including structure prediction using the adaptive genetic algorithm (AGA) (Wu et al., 2014). Section 3 presents and discusses the discovered high-pressure phases, assessing their thermodynamic stability at elevated temperatures. We conclude this study by summarizing our findings in Section 4.

## 2. Computational methods

First-principles calculations were performed using the finite-temperature version of density functional theory (DFT) (Wentzcovitch et al., 1992; Mermin et al., 1965), as implemented in the Quantum ESPRESSO (QE) package (Giannozzi et al., 2009; Giannozzi et al., 2017). The exchange-correlation functional was treated using the non-spin-polarized local density approximation (LDA). Spin-polarized calculations were performed for the low-energy phases, but the magnetic moments were found to be quenched in self-consistent calculations. Vanderbilt-type ultrasoft pseudopotentials (Vanderbilt et al., 1990) were employed for Fe, Si, and O, with the valence electronic configuration of $3s^2 3p^6 3d^{6.5} 4s^1$ (Fe), $3s^2 3p^1$ (Si), and $2s^2 2p^4$ (O). These potentials were previously tested and used in a few studies at TPa pressures (Zheng et al., 2022; Umemoto et al., 2008). A kinetic energy cutoff of 50 Ry and a charge density cutoff of 500 Ry were applied. Brillouin-zone integration was performed over a Monkhorst-Pack $k$-point grid of $2\pi \times 0.03$ Å$^{-1}$ in the structure optimization. Structural optimization was performed under constant pressure using the Broyden-Fletcher Goldfarb-Shanno (BFGS) algorithm (Broyden et al., 1970; Fletcher et al., 1970; Goldfarb et al., 1970; Shanno et al., 1985) with variable cell shape. The convergence thresholds are 0.01 eV/Å for the atomic force, 0.01 GPa for stress tensor components, and $1 \times 10^{-5}$ eV for the total energy.

The finite displacement method was used to compute phonon frequencies. Supercells for phonon calculations were (4×4×3) for $P3$ $FeSiO_4$, (2×2×2) for $P3$ $Fe_4Si_5O_{18}$, and (2×2×4) for $P\overline{3}$ $FeSi_2O_6$, with a uniform Monkhorst-Pack k-point mesh of density $2\pi \times 0.03$ Å$^{-1}$. The second-order force constants and harmonic phonon dispersion curves were obtained using the PHONOPY code (Togo et al., 2015). The thermodynamic properties at finite temperatures were investigated within the quasi-harmonic approximation (QHA) (Althoff et al., 1993; Wallace et al., 1972) and the qha code (Qin et al., 2019). Although the electronic entropy affects phonons, we used the electronic entropy at 4000 K to approximate the contributions at all temperatures.

For crystal structure prediction using the AGA (Wu et al., 2014) assisted by classical auxiliary embedded-atom method (EAM) potentials (Foiles et al., 1986) was employed**.** In EAM, the total energy of an $N$-atom system was evaluated by

$$E_{\text{total}} = \frac{1}{2}\sum_{i,j(i\neq j)}^{N} \phi(r_{ij}) + \sum_{i} F_i(n_i), \quad (1)$$

where $\phi(r_{ij})$ denotes the pair repulsion between atoms $i$ and $j$ with a distance of $r_{ij}$; $F_i(n_i)$ is the embedded term with the electron density term $n_i = \sum_{j\neq i} \rho_j\left(r_{ij}\right)$ at the site occupied by atom $i$. The fitting parameters in the EAM formula were chosen as follows: The pair interactions were modeled by the Morse function,

$$\phi(r_{ij}) = D\left[e^{-2\alpha(r_{ij}-r_0)} - 2e^{-\alpha(r_{ij}-r_0)}\right], \quad (2)$$

where $D$, $\alpha$ and $r_0$ are fitting parameters. The density function is modeled by an exponentially decaying function,

$$\rho(r_{ij}) = \alpha \exp[-\beta(r_{ij} - r_0)], \quad (3)$$

where $\alpha$ and $\beta$ are fitting parameters. The form proposed by Banerjee and Smith in Ref. (Banerjea et al., 1988) was used as the embedding function with fitting parameters $F_0$, $\gamma$ as

$$F(n) = F_0[1 - \gamma \ln n]n^{\gamma}. \quad (4)$$

The potential fitting was performed using the force-matching method with a stochastic simulated annealing algorithm as implemented in the POTFIT code (Brommer et al., 2006; Brommer et al., 2007).

## 3. Results and discussion

### 3.1 Survey of relevant binary systems

The phase transitions of the elemental and binary phases of the Fe-Si-O system have been extensively studied in previous works. We first review the ground-state structures of the most stable binaries $SiO_2$ (Wu et al., 2011; Tsuchiya et al., 2011; Liu et al., 2021), $FeO_2$, $Fe_2O$, and FeO (Zheng et al., 2022; Weerasinghe et al., 2015) within the 0.5-1.3 TPa pressure range (Fig. 1(a)). Their crystal structures are shown in Fig. 1(b) and (c), where the high-temperature phase Cotunnite-type $SiO_2$ is also included. The pyrite-type phase ($Pa\bar{3}$) of $SiO_2$ transitions to the $R\bar{3}$ phase at 667 GPa, which then transitions to the $Fe_2P$-type structure above 914 GPa. $FeO_2$ transitions from $R\bar{3}m$ symmetry to the orthorhombic *Pnma* structure at 562 GPa, and this *Pnma* phase remains stable up to 1.3 Tpa. $Fe_2O$ retains its *I*4/*mmm* structure throughout the entire pressure range. FeO transitions from the $R\bar{3}$ phase to *Cmcm* symmetry at ~570 GPa, and then to the B2 (CsCl-type) structure above 835 GPa. These phase transitions are consistent with the work of others (Wu et al., 2011; Tsuchiya et al., 2011; Liu et al., 2021; Zheng et al., 2022; Weerasinghe et al., 2015), although minor discrepancies in transition pressures may stem from differences in computational methods.

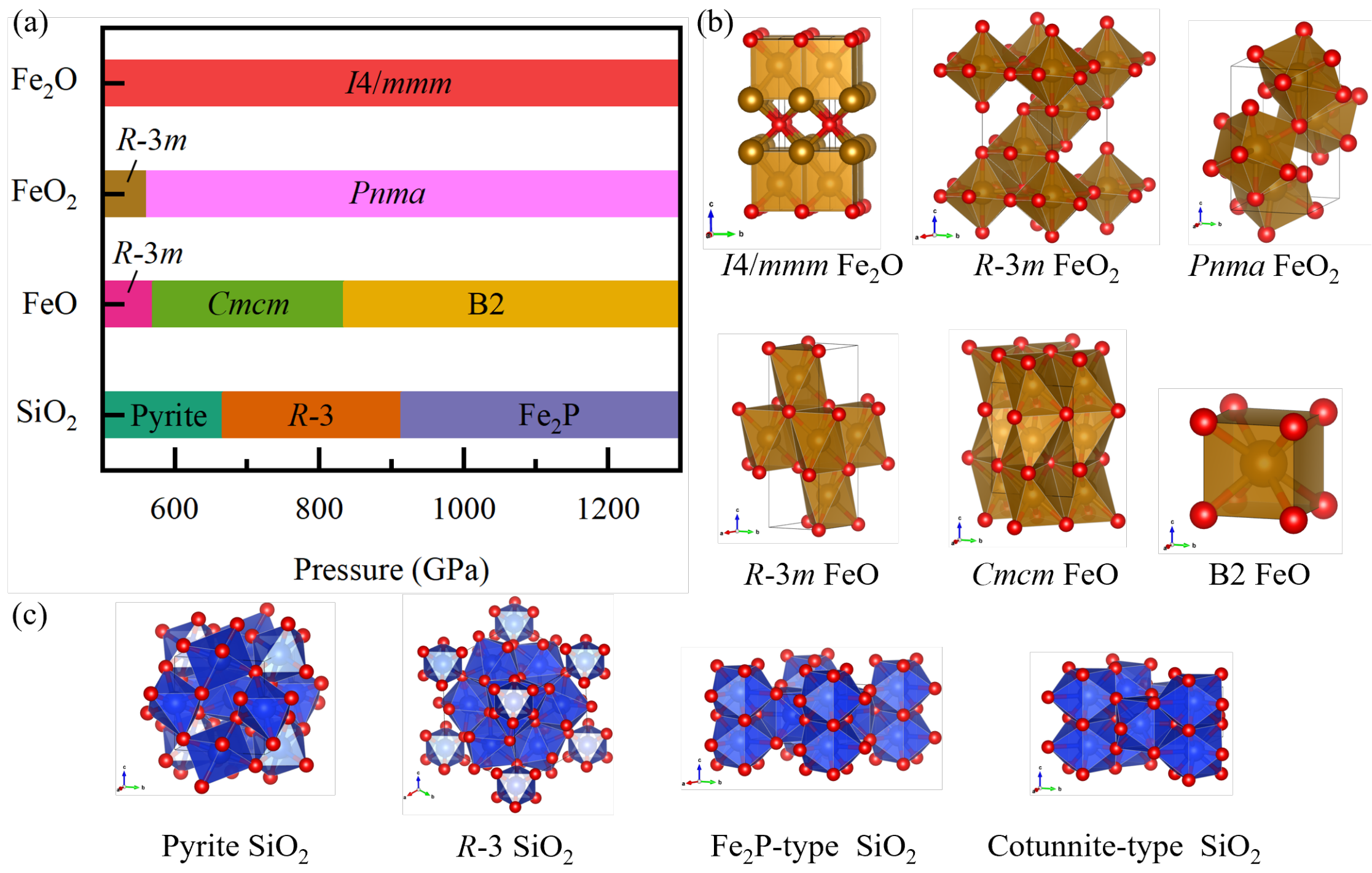

**Figure 1.** (a) Phase relations in $SiO_2$, $FeO_2$, $Fe_2O$, and FeO within the 0.5–1.3 Tpa pressure range. Crystal structures of (b) Fe-O binaries and (c) $SiO_2$ mentioned in (a). The Cotunnite-type in (c) is a high-temperature version $Fe_2P$-type $SiO_2$.

### 3.2 Crystal structure prediction

Using our AGA crystal structure prediction method (Wu et al., 2014), we performed successive structural searches for the Fe-Si-O system at 1 Tpa. We systematically explored the compositional space of the Fe-Si-O system by selecting distinct components characterized by the formula $Fe_xSi_yO_z$. The parameters x, y, and z are integers constrained to the set $\{(x,y,z) \mid x,y,z \in \{1,2,3,4\} \text{ and } x+y+z \leqslant 10\}$. Each combination was reduced to its simplest form and used for the search, with 1 to 10 formula units and no more than 30 atoms per cell. We have also explored prospective pseudo-binaries of the Fe-Si-O system. By selecting multiple pairs of edge end members in the convex hull, we have conducted searches by combining them in the ratios of 1:1, 1:2, 1:3, 2:1, and 3:1, respectively.

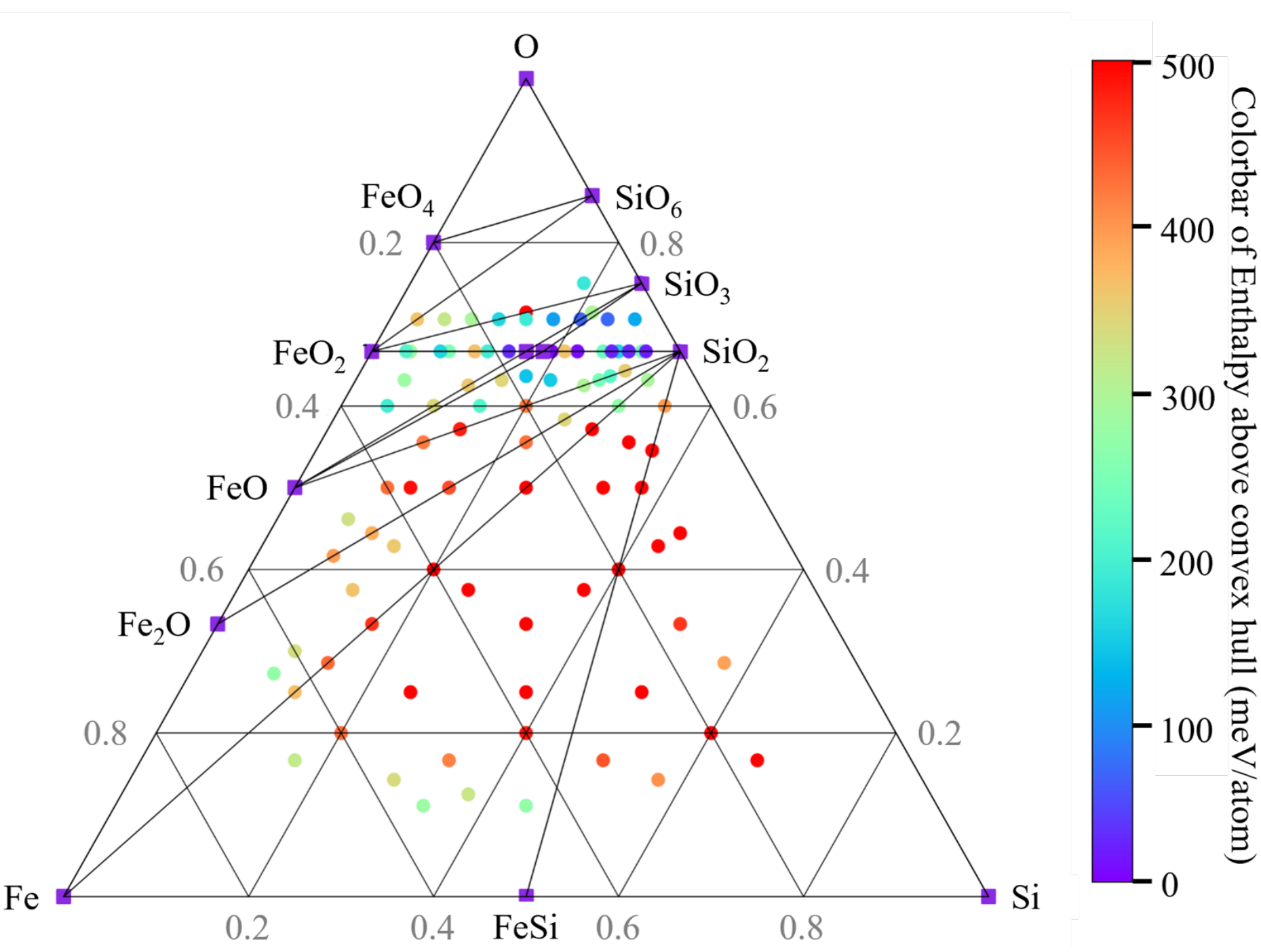

**Figure 2.** Convex hull of the Fe-Si-O system at 1 Tpa and 0 K. Purple squares indicate endmembers and compounds on the convex hull, and solid circles represent the compositions searched in this study. The color gradient from purple to red signifies an increasing magnitude of $H_d$ (formation enthalpy distance above the convex hull.

After a series of searches to locate energetically favorable stoichiometries, refining the search, and performing configuration substitutions, three stable phases were identified: $P3$ $FeSiO_4$, $P3$ $Fe_4Si_5O_{18}$, and $P\bar{3}$ $FeSi_2O_6$, among which only $P\bar{3}$ $FeSi_2O_6$ exhibited thermodynamic stability at high temperatures. The absence of imaginary vibrational frequencies indicates good dynamical stability of these phases (Fig. S3). The band structures and the projected electronic density of states (Fig. S3) show that all three structures are metallic at 1 Tpa, with the electronic density of states dominated by oxygen and iron states.

Fig. 2 illustrates the convex hull at 1 TPa and 0 K, where phases with formation enthalpies ($\Delta H$) below the convex hull are thermodynamically stable. The stability hierarchy is quantified by the distance to the convex hull, $H_d$, where $H_d = 0$ corresponds to the ground state.

### 3.3 Structural analysis

The crystal structures of $P3$ $FeSiO_4$, $P3$ $Fe_4Si_5O_{18}$ and $P\overline{3}$ $FeSi_2O_6$ are shown in Fig. 3(a). These crystalline phases exhibit a high degree of structural similarity. They bear a marked resemblance to the $Fe_2P$-type $SiO_2$, possessing similar Si-O polyhedra. The non-oxygen ions within the *ab*-plane have identical arrangements. Notably, Fe in $P3$ $FeSiO_4$ and $P3$ $Fe_4Si_5O_{18}$ exhibit two distinct coordination environments: six-fold and nine-fold. In contrast, only six-fold coordinated Fe ions are found in the $P\overline{3}$ $FeSi_2O_6$ phase. Six-fold ions are configured in an octahedral geometry, while the nine-fold sites form tri-capped trigonal prisms ($XO_6$) augmented by three additional oxygen ions ($XO_{6+3}$) positioned outside the centers of the rectangular faces, as indicated by the white and black lines in Fig. 3(b). The oxygen sublattice in the three structures, as well as in $Fe_2P$-type $SiO_2$, is composed of distorted hexagonal layers that are stacked along the *c* axis and exhibit relative slip (displacement) between adjacent layers parallel to the *ab*-plane. Considering the oxygen sublattice as the primary framework, all these structures can be conceptualized as arising from the filling of interstitial sites within it. The nine-fold silicon or iron co-plane is parallel to the oxygen layers, whereas the six-fold iron intercalates between them. Let the height of a single tri-capped prism in the *c*-direction be denoted as *d*. Polyhedra with the same coordination stack along the c-axis. The tri-capped prisms are connected via sharing their apical (or basal) faces. Adjacent columns exhibit a relative displacement of $d/2$ along the *c*-axis; correspondingly, two octahedra within the same column are also separated by a distance of *d*. Oxygens are exclusively four-fold coordinated with iron or silicon, forming tetrahedral structures. Only two distinct coordination motifs are observed in these tetrahedra: O-$Fe_2Si_2$ and O-$FeSi_3$.

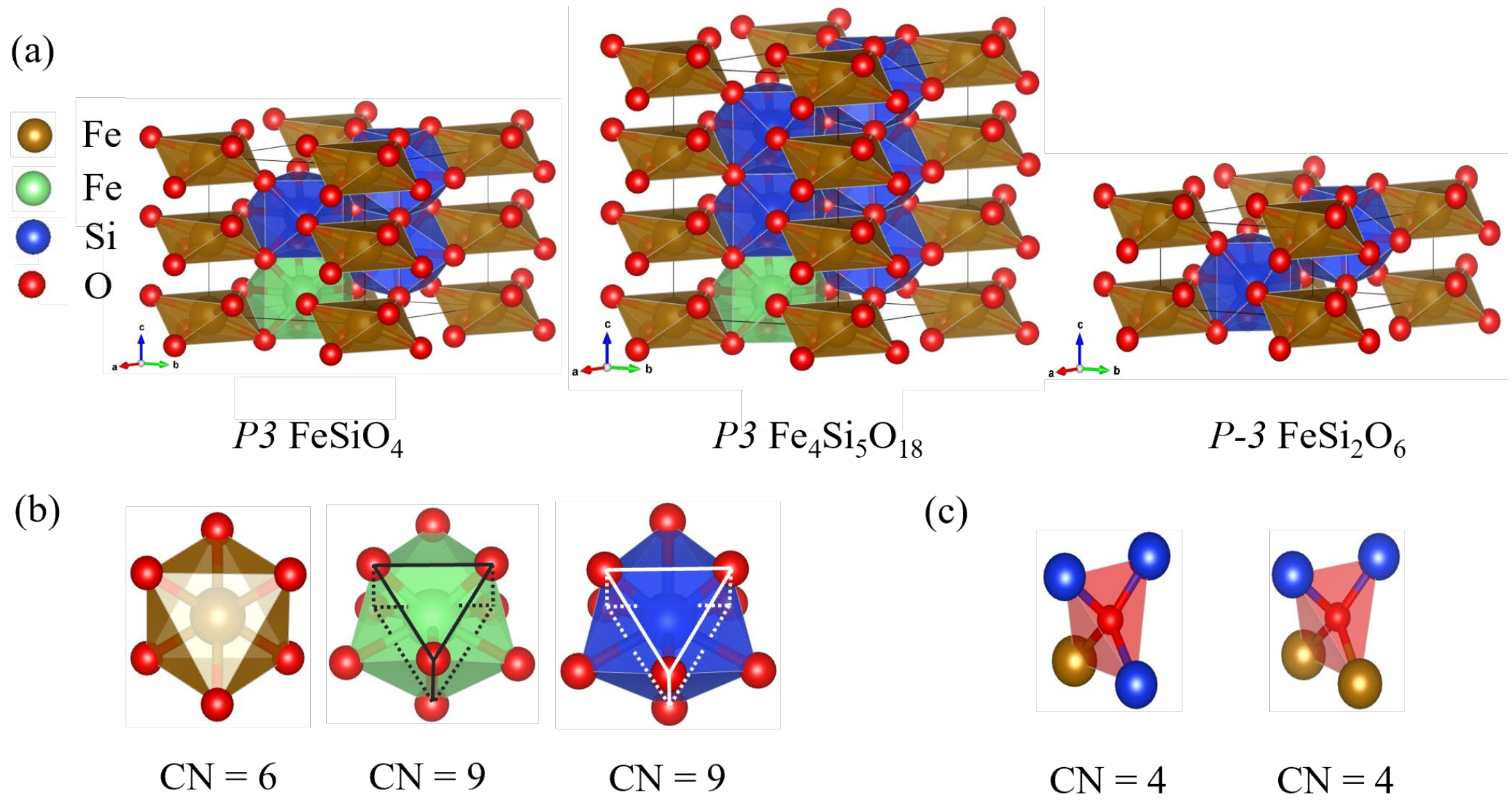

**Figure 3.** Structural details of novel Fe-Si-O ternary phases. (a) Crystal structures of *P*3 $FeSiO_4$ (left), *P*3 $Fe_4Si_5O_{18}$ (middle), and $P\overline{3}$ $FeSi_2O_6$ (right). Atoms are represented by spheres of different colors: brown for iron (Fe), blue for silicon (Si), and red for oxygen (O). (b) Three polyhedral types, i.e., $FeO_6$ (brown), $FeO_9$ (green), and $SiO_9$ (blue) exist in these structures. The white and black lines indicate the trigonal prisms centered on non-oxygen ions. (c) Polyhedra centered on oxygens are four-fold coordinated and form tetrahedra.

Structural comparison reveals a series of analogous compounds to these novel Fe-Si-O ternary phases, namely $NaLnF_4$ (Ln = La, Ce, Pr, Nd, Sm, or Gd) (Grzechnik et al., 2012; Shi et al., 2021; Grzechnik et al., 2002). These compounds exhibit intrinsic disorder, with the six-fold coordinated sites occupied by Na, while the nine-fold coordinated sites are occupied by a mixture of Na and Ln. Another series of halides has also been found to possess structures similar to our newly discovered structures (Mao et al., 2025). The existence of these low-pressure analogs allows for an effective study of the properties and structural evolution of the Fe-Si-O high-pressure system under laboratory conditions, i.e., whether dissociation into binary occurs under pressure and what the binary structures would be, similar to the low-pressure analogs of $MgSiO_3$, namely, $NaMgF_3$ and $MgGeO_3$ (Umemoto et al., 2019; Umemoto et al., 2021).

It is noted that *P*3 $FeSiO_4$ and *P*3 $Fe_4Si_5O_{18}$ can be derived through the stacking of $P\overline{3}$ $FeSi_2O_6$ units and atomic Si-Fe replacements (Fig. 3(a)). Consequently, it is reasonable to regard $P\overline{3}$ $FeSi_2O_6$ is the basic unit for this category of such pseudo-binaries. Here, we aim to elucidate the structural relationship between $P\overline{3}$ $FeSi_2O_6$ and the endmembers of this pseudo-binary system ($Fe_2P$-type $SiO_2$ and *Pnma* $FeO_2$).

Starting from the $Fe_2P$-type $SiO_2$ structure, substituting one third Si ions in the primitive cell with Fe and performing structural optimization yields a $P\overline{6}$ $FeSi_2O_6$ structure (preserving the $Fe_2P$-type configuration). However, the corresponding phonon dispersion indicates that this phase is dynamically unstable (Fig. S4(c)). By performing structural relaxation guided by the displacement of the most unstable eigenmode at the Γ-point of the Brillouin zone, the system ultimately relaxed into the stable $P\overline{3}$ $FeSi_2O_6$ phase.

Similarly, starting from *Pnma* $FeO_2$, replacing all Fe with Si yields cotunnite-type $SiO_2$ after relaxation. Recall that the cotunnite-type $SiO_2$ is the high-temperature version of $Fe_2P$-type $SiO_2$, with a well-known structural relation between them (Wu et al., 2011). Based on this relationship, the primitive cell of cotunnite-type $SiO_2$ was expanded threefold along the c-axis, followed by site-specific Fe-Si substitutions to match the stoichiometry and Fe-Si ordering of $P\overline{3}$ $FeSi_2O_6$ (Fig. S4(b)). A structural optimization of this configuration leads to *Pnma* $FeSi_2O_6$ (distorted cotunnite-type structure). Upon introducing a shift of four Si atoms along the b-axis, as in $SiO_2$ (Wu et al., 2011), a secondary relaxation results in the transition to the $P2_1/m$ $FeSi_2O_6$ phase (Fig. 4(a)). Our calculations indicate that this process involves an energy barrier of 0.32 eV/atom (~3700 K, Fig. S4(d)). Nevertheless, the extreme conditions of super-Earth mantles render this barrier surmountable. Employing a similar procedure to eliminate the phonon instabilities (Fig. S4(c)), we ultimately attained an orthorhombic supercell of the $P\overline{3}$ phase.

The fact that both binary endmembers can yield the $P\overline{3}$ $FeSi_2O_6$ phase by targeted ionic substitutions and structural relaxations, provides compelling evidence that $P\overline{3}$ $FeSi_2O_6$ is a well-defined pseudo-binary phase whose structure is closely related to those of the endmembers $FeO_2$ and $SiO_2$.

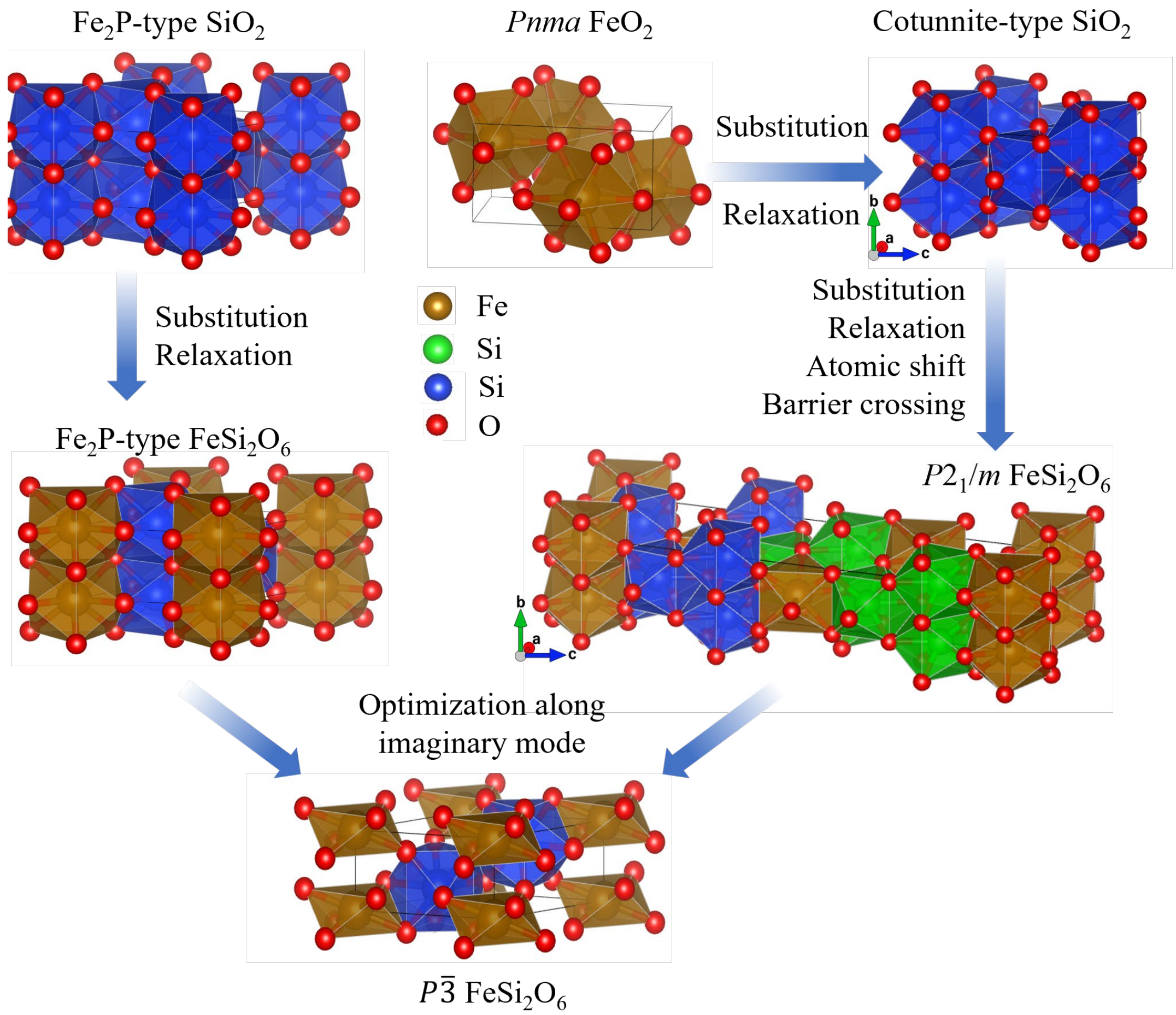

**Figure 4.** Two pathways from $Fe_2P$-type $SiO_2$ and *Pnma* $FeO_2$ at 1 TPa to $P\bar{3}$ $FeSi_2O_6$. The green Si atoms shown in are required to be shifted along the *b*-axis.

### 3.4 Energetics under Pressure and Increased Temperature

To investigate the high-temperature stability of these phases, we computed the Gibbs free energy of all phases involved and the phase diagram for the $FeO_2$-$SiO_2$ pseudo-binary system at 1 TPa. As depicted in Fig. 5(a), with increasing temperature, the *P*3 $FeSiO_4$ and *P*3 $Fe_4Si_5O_{18}$ phases transition from stable to metastable states. Conversely, $P\bar{3}$ $FeSi_2O_6$ emerges as the sole stable phase within the pseudo-binary system above 4000 K. Fig. 5(b) shows free energies excluding vibrational contributions. The inclusion of vibrational contributions further stabilizes all three ternary phases relative to their binary end members. The considerably higher density of states at the Fermi level in $P\overline{3}$ $FeSi_2O_6$ makes the electronic entropy a major stabilization factor for this phase at high temperatures (Fig. S3(c)).

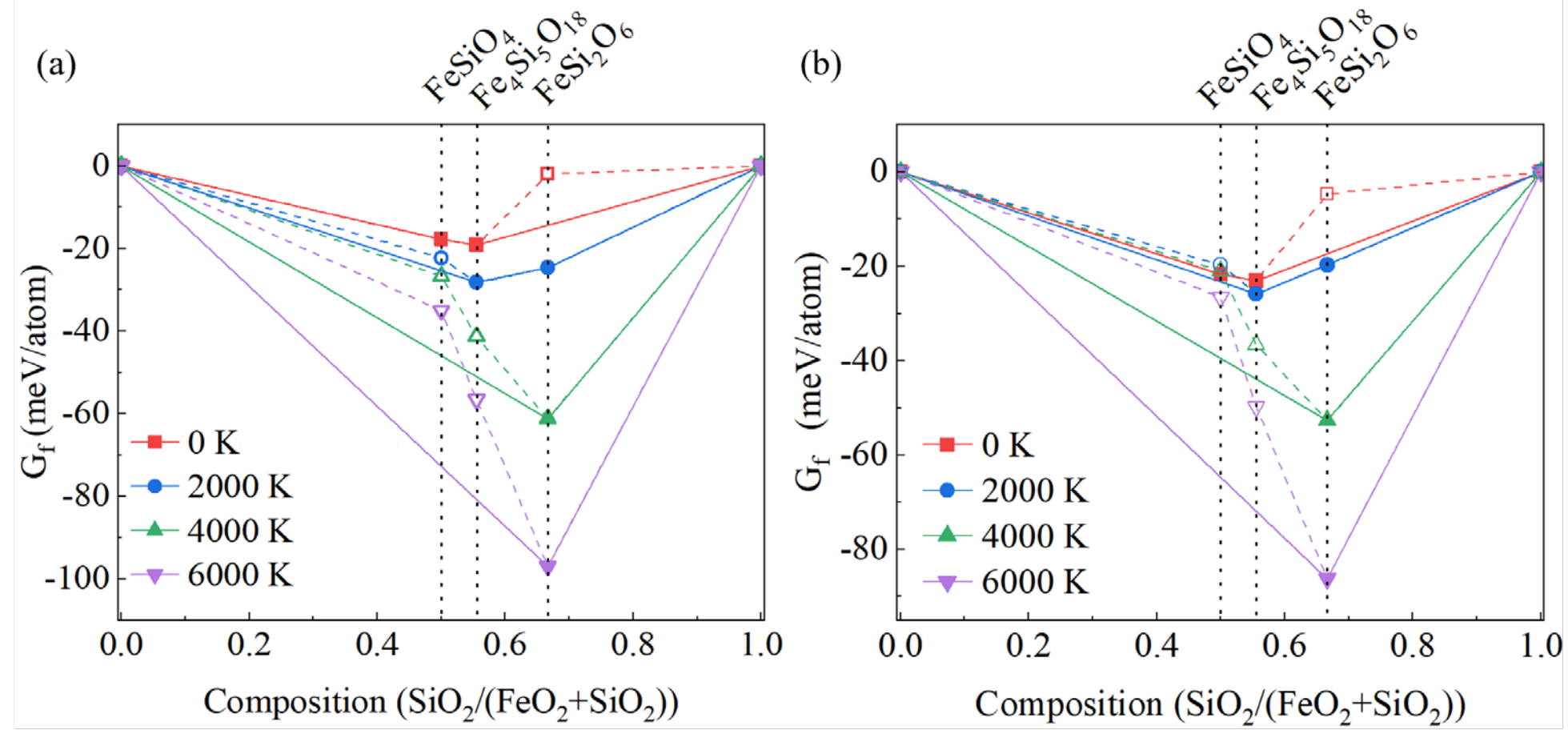

**Figure 5.** (a) Gibbs free energies with increasing temperature, including vibrational and electronic excitation contributions at 1 tPa. (b) Free energies including only thermal electronic excitation. Free energies are plotted relative to the ground state phases of $FeO_2$ and $SiO_2$ at various temperatures. Solid markers represent convex-hull points, while open markers indicate metastable structures.

We evaluated the stability fields of the three phases identified in the Fe-Si-O system at Fe/Si ratios of 1.0, 0.8, and 0.5. Fig. 6 presents the pressure-temperature conditions at the core-mantle boundary (CMB) of super-Earths (van den Berg et al., 2019) with several $M_\oplus$. It shows that the Fe-Si-O system exhibits complex phase transitions within the temperature and pressure ranges investigated in our study.

For Fe/Si = 1.0, $FeSiO_4$ is stable in terrestrial mantles of super-Earths with up to ~4.5$M_\oplus$. In larger planets, $FeSiO_4$ dissociates into $FeSi_2O_6$ and $FeO_2$ (see Fig. 6(a)). For 1/2 < Fe/Si < 1, e.g., 4/5, $FeSiO_4$ and $SiO_2$ coexist for masses below ~3.5$M_\oplus$. Between ~3.5 and ~4.5$M_\oplus$ the system undergoes a recombination transition in which $FeSiO_4$ plus $FeSi_2O_6$ coexist, whereas for masses in excess of ~4.5$M_\oplus$ $FeSi_2O_6$ and $FeO_2$ coexist in equilibrium (Fig. 6(b)). For Fe/Si ≤ 1/2, e.g., 1/2, $FeSiO_4$ plus $SiO_2$ are stable for masses up to ~3.5$M_\oplus$, with $FeSi_2O_6$ stabilizing in mantles of planets more massive than that (Fig. 6(c)).

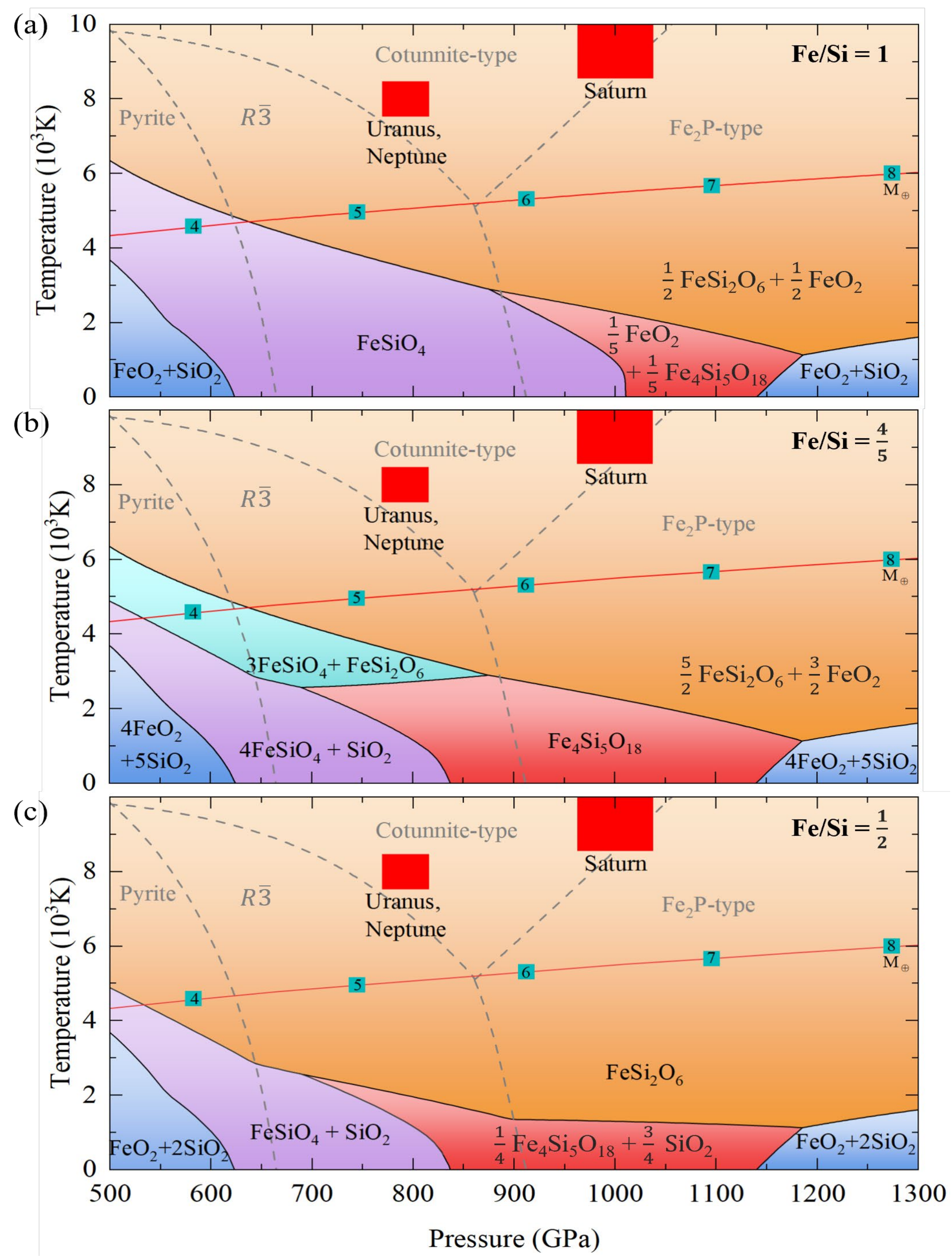

**Figure 6.** Phase diagram of the three new phases: (a) *P*3 $FeSiO_4$, (b) *P*3 $Fe_4Si_5O_{18}$, and (c) $P\bar{3}$ $FeSi_2O_6$. The red line and small blue squares show the CMB pressure-temperature conditions for super-Earths (van den Berg et al., 2019). Associated numbers refer to the planet mass expressed in Earth-mass units ($M_\oplus$). Red squares mark Guillot's estimated CMB pressure-temperature conditions for the solar giants (Guillot et al., 2004). The gray dashed lines mark the phase boundaries of $SiO_2$.

At super-Earth CMB conditions, with increasing depth, the fate of *P*3 $FeSiO_4$ is controlled by the local $SiO_2$ budget. In $SiO_2$-deficient environments (e.g., Fe/Si = 1), $FeSiO_4$ dissociates into $FeSi_2O_6$ and $FeO_2$ (Clapeyron slope (CS) = −109.9 mPa/K); when $SiO_2$ is slightly in excess (Fe/Si = 4/5), $FeSiO_4$ efficiently scavenges the available $SiO_2$ to form $FeSi_2O_6$ (CS = −72.2 mPa/K), while any residual $FeSiO_4$ is subsequently converted to $FeSi_2O_6$ plus $FeO_2$ (CS = −109.9 mPa/K); in $SiO_2$-rich regimes (Fe/Si = 1/2), all $FeSiO_4$ can recombine with $SiO_2$ to produce $FeSi_2O_6$ (CS = −72.2 mPa/K). These CSs are unusually large compared to phase transitions in the Earth's mantle. Phase transitions with negative CS inhibit flow across such boundary layers. These very large CSs seen here might induce layering in Super-Earths' mantles (Tackley et al., 1993).

## 4. Summary and conclusions

We systematically investigated high-pressure phase relations in the Fe-Si-O system at ~ 1 TPa using the AGA method combined with *ab initio* LDA calculations. Two stoichiometric compounds, *P*3 $FeSiO_4$ and *P*3 $Fe_4Si_5O_{18}$, were found to be stable at 0K, while $P\bar{3}$ $FeSi_2O_6$ was shown to stabilize at high temperatures. These compounds exhibit similar structural building blocks (polyhedra) and can be viewed as $FeO_2$-$SiO_2$ pseudo-binary phases characterized as metallic, paramagnetic systems.

The crystal structure of $P\bar{3}$ $FeSi_2O_6$ can be traced to the structures of the stable elementary oxides at ~1 TPa, namely 1 TPa, i.e., $Fe_2P$-type $SiO_2$ and *Pnma*-type $FeO_2$. The $P\bar{3}$ $FeSi_2O_6$ structure can be generated by appropriate substitution of Si with Fe in $Fe_2P$-type $SiO_2$ followed by a structural relaxation. Its relationship with *Pnma*-type $FeO_2$ is more subtle. Replacing all Fe with Si in *Pnma*-type $FeO_2$ yields cotunnite-type $SiO_2$, the high-temperature polymorph of $Fe_2P$-type $SiO_2$. Partial substitution of Fe with Si in *Pnma*-type $FeO_2$ produces $P2_1/m$ $FeSi_2O_6$, which is related to $P\bar{3}$ $FeSi_2O_6$ in the same way that cotunnite-type $SiO_2$ is related to $Fe_2P$-type $SiO_2$. The related phases *P*3 $FeSiO_4$ and *P*3 $Fe_4Si_5O_{18}$ can be derived directly from $P\bar{3}$$FeSi_2O_6$ through further targeted substitution of Si by Fe.

In the Earth's silicate mantle, $Fe^{+2}$ enters as $[Fe]_{Mg}$ in Mg-silicate phases, e.g., $(Mg,Fe)SiO_3$ bridgmanite and $(Mg,Fe)_2SiO_4$ olivine. $Fe^{+3}$ can enter as a coupled substitution $[2Fe]_{Mg,Si}$, e.g., $(Mg,Fe)(Si,Fe)O_3$ in bridgmanite, but usually, aluminum is favored on the silicon site ($[Al]_{Si}$) (Hsu et al., 2012). By analogy, Fe is also thought to "dissolve" mainly as $[Fe]_{Mg}$ or $[2Fe]_{Mg,Si}$ in the canonical TPa silicate hosts, i.e., $I\bar{4}2d$-type $Mg_2SiO_4$ or $P2_1/c$-type $MgSi_2O_5$. Prior to this work, no stable Fe-Si-O compounds have ever been reported at TPa pressures. This discovery of the $(Fe,Si)O_2$ pseudo-binary phases may not necessarily alter the phase transition sequence of the Mg-Si-O system if Fe is present in low concentrations. It is expected, however, that small amounts of Fe will open two-phase fields (binary loops) in the Mg-Si-O phase diagram (Umemoto et al., 2017).

The newly pseudo-binaries found here show that Fe favors an $FeO_2$-type component over FeO, $Fe_2SiO_4$, and $Fe_2O_3$, and possibly *Pnma*-type $FeMgO_3$ or *Im-3m*-type $FeMg_3O_4$ at TPa pressures (Fang et al., 2023), the latter ones yet to be investigated in the quaternary Mg-Si-Fe-O context. Mg/Si/Fe disorder in $I\bar{4}2d$-type $Mg_2SiO_4$ (Umemoto et al., 2021; Dutta et al., 2023) might help stabilize iron in the $I\bar{4}2d$-type silicate, and a careful study of the disordered $(Mg,Si,Fe)_2(Mg,Si,Fe)O_4$ system should be carried out. It is also possible that the stability of these phases may trigger silicate dissociation into oxides ($(Mg,Fe)_2(Si,Fe)O_4 \rightarrow 2(Mg,Fe)O + (Si,Fe)O_2$) below ~3 TPa (Umemoto et al., 2017), with the extent of the dissociation governed by the iron content.

In short, these newly discovered $FeO_2$-$SiO_2$ pseudo-binaries imply a markedly different pattern of Fe incorporation in mantle phases at TPa conditions, compared with the behavior inferred at the GPa pressures of Earth's mantle.

## Acknowledgments

This research was supported by the National Natural Science Foundation of China (Nos. 12374015, T2422016, and 42374108). Shaorong Fang and Tianfu Wu from the Information and Network Center at Xiamen University are acknowledged for their assistance with Graphics Processing Unit (GPU) computing. The work at Columbia University was supported by the Gordon and Betty Moore Foundation Award GBMF12801 42 (doi.org/10.37807/GBMF12801) and results from a collaboration established under NSF Grants EAR-2000850 and EAR-1918126.

## References

Althoff, J.D., Allen, P.B., Wentzcovitch, R.M., Moriarty, J.A., 1993. Phase diagram and thermodynamic properties of solid magnesium in the quasiharmonic approximation. Phys. Rev. B 48, 13253.

Banerjea, A., Smith, J.R., 1988. Origins of the universal binding-energy relation. Phys. Rev. B 37, 6632.

Brommer, P., Gähler, F., 2006. Effective potentials for quasicrystals from ab-initio data. Philos. Mag. 86, 753.

Brommer, P., Gähler, F., 2007. Potfit: effective potentials from ab initio data. Modell. Simul. Mater. Sci. Eng. 15, 295.

Broyden, C.G., 1970. The Convergence of a Class of Double-rank Minimization Algorithms 2. The New Algorithm. IMA J. Appl. Math. 6, 222.

Dutta, R., Greenberg, E., Prakapenka, V.B., et al., 2019. Phase Transitions beyond post-perovskite in $NaMgF_3$ to 160 GPa. Proc. Nat. Acad. Sci. 116, 19324–19329.

Dutta, R., Tracy, S.J., Cohen, R.E., et al., 2022. Ultrahigh-pressure disordered eight-coordinated phase of $Mg_2GeO_4$: Analogue for super-Earth mantles. Proc. Natl. Acad. Sci. U.S.A. 119, e2114424119.

Dutta, R., Tracy, S.J., Cohen, R.E., 2023. High-pressure order-disorder transition in $Mg_2SiO_4$: Implications for super-Earth mineralogy. Phys. Rev. B 107, 184112.

Fang, Y.M., Sun, Y., Wang, R.H., et al., 2023. Structural prediction of Fe-Mg-O

compounds at super-Earth's pressures. Phys. Rev. Materials 7, 113602.

Fletcher, R., 1970. A new approach to variable metric algorithms. Comput. J. 13, 317.

Foiles, S.M., Baskes, M.I., Daw, M.S., 1986. Embedded-atom-method functions for the fcc metals Cu, Ag, Au, Ni, Pd, Pt, and their alloys. Phys. Rev. B 33, 7983.

Fortney, J.J., Marley, M.S., Barnes, J.W., 2007. Planetary Radii across Five Orders of Magnitude in Mass and Stellar Insolation: Application to Transits. Astrophys. J. 659, 1661.

Giannozzi, P., et al., 2009. QUANTUM ESPRESSO: a modular and open-source software project for quantum simulations of materials. J. Phys.: Condens. Matter 21 395502.

Giannozzi, P., et al., 2017. Advanced capabilities for materials modelling with QUANTUM ESPRESSO. J. Phys.: Condens. Matter 29 465901.

Goldfarb, D., 1970. A family of variable-metric methods derived by variational means. Math. Comput. 24, 23.

Grocholski, B., Shim, S.-H., Prakapenka, V.B., 2010. Stability of the $MgSiO_3$ analog $NaMgF_3$ and its implication for mantle structure in super-Earths. Geophys. Res. Lett. 37, L14204.

Grzechnik, A., Friese, K., 2012. Crystal structures and stability of $NaLnF_4$ (Ln = La, Ce, Pr, Nd, Sm and Gd) studied with synchrotron single-crystal and powder diffraction. Dalton Trans. 41, 10258.

Grzechnik, A., Bouvier, P., Mezouar, M., et al., 2002. Hexagonal $Na_{1.5}Y_{1.5}F_6$ at high pressures. J. Solid State Chem. 165, 159–164.

Guillot, T., 2004. Special issue: Probing the giant planets. Phys. Today 57, 63.

Henkelman, G., Jónsson, H., 2000. Improved tangent estimate in the nudged elastic band method for finding minimum energy paths and saddle points. J. Chem. Phys. 113, 9978–9985.

Hsu, H., Yu, Y.G., Wentzcovitch, R.M., 2012. Spin crossover of iron in aluminous MgSiO3 perovskite and post-perovskite. Earth Planet. Sci. Lett. 359, 34–39.

Liu, C., et al., 2021. Mixed Coordination Silica at Megabar Pressure. Phys Rev Lett.

126, 035701.

Mao, X., Dawson, J.A., 2025. Optimizing Li-Ion Transport in $LaCl_{3-x}Br_x$ solid electrolytes through anion mixing. EcoMat 7, e70006.

McArthur, B.E., Endl, M., Cochran, W.D., et al., 2004. Detection of a Neptune-Mass Planet in the ρ1 Cancri System Using the Hobby-Eberly Telescope. Astrophys. J. 614, L81.

Mermin, N.D., 1965. Thermal Properties of the Inhomogeneous Electron Gas. Phys. Rev. 137, A1441.

Murakami, M., Hirose, K., Kawamura, K., Sata, N., Ohishi, Y., 2004. Post-Perovskite Phase Transition in $MgSiO_3$. Science 304, 855–858.

Niu, H., Oganov, A.R., Chen, X.Q., Li, D., 2016. Prediction of novel stable compounds in the Mg–Si–O system under exoplanet pressures. Sci Rep 5, 18347.

Oganov, A.R., Ono, S., 2004. Theoretical and experimental evidence for a post-perovskite phase of $MgSiO_3$ in Earth's D" layer. Nature 430, 445–448.

Qin, T., Zhang, Q., Wentzcovitch, R.M., Umemoto, K., 2019. qha: A Python package for quasiharmonic free energy calculation for multi-configuration systems. Comput. Phys. Commun. 237, 199.

Seager, S., Kuchner, M., Hier-Majumder, C.A., et al., 2007. Mass-Radius Relationships for Solid Exoplanets. Astrophys. J. 669, 1279.

Shanno, D.F., 1985. An example of numerical nonconvergence of a variable-metric method. Math. Comput. 24, 647.

Shi, R., Brites, C.D.S., Carlos, L.D., 2021. Hexagonal-phase $NaREF_4$ upconversion nanocrystals: the matter of crystal structure. Nanoscale 13, 19771.

Tackley, P.J., Stevenson, D.J., Glatzmaier, G.A., et al., 1993. Effects of an endothermic phase transition at 670 km depth in a spherical model of convection in the Earth's mantle. Nature 361, 699–704.

Togo, A., Tanaka, I., 2015. First principles phonon calculations in materials science. Scripta Mater. 108, 1.

Tsuchiya, T., Tsuchiya, J., Umemoto, K., Wentzcovitch, R.M., 2004. Phase transition in $MgSiO_3$ perovskite in the Earth's lower mantle. Earth Planet. Sci. Lett. 224, 241–248.

Tsuchiya, T., Tsuchiya, J., 2011. Prediction of a hexagonal $SiO_2$ phase affecting stabilities of $MgSiO_3$ and $CaSiO_3$ at multimegabar pressures. Proc. Natl. Acad. Sci. U.S.A. 108, 1252.

Umemoto, K., Wentzcovitch, R.M., Wu, S.Q., et al., 2017. Phase transitions in $MgSiO_3$ post-perovskite in super-Earth mantles. Earth Planet. Sci. Lett. 478, 40–45.

Umemoto, K., Wentzcovitch, R.M., Allen, P.B., 2006. Dissociation of $MgSiO_3$ in the Cores of Gas Giants and Terrestrial Exoplanets. Science 311, 983–986.

Umemoto, K., Wentzcovitch, R.M., 2011. Two-stage dissociation in $MgSiO_3$ post-perovskite. Earth Planet. Sci. Lett. 311, 225–229.

Umemoto, K., Wentzcovitch, R.M., 2019. Ab initio exploration of post-PPV transitions in low-pressure analogs of $MgSiO_3$. Phys. Rev. Materials 3, 123601.

Umemoto, K., Wentzcovitch, R.M., 2021. Ab initio prediction of an order-disorder transition in $Mg_2GeO_4$: Implication for the nature of super-Earth's mantles. Phys. Rev. Materials 5, 093604.

Umemoto, K., Wentzcovitch, R.M., Yu, Y.G., Requist, R., 2008. Spin transition in $(Mg,Fe)SiO_3$ perovskite under pressure. Earth Planet. Sci. Lett. 276, 198.

van den Berg, A.P., Umemoto, K., Yuen, D., Wentzcovitch, R.M., 2019. Mass-dependent dynamics of terrestrial exoplanets using ab initio mineral properties. Icarus 317, 412–426.

Vanderbilt, D., 1990. Soft self-consistent pseudopotentials in a generalized eigenvalue formalism. Phys. Rev. B 41, 7892.

Wallace, D.C., Callen, H., 1972. Thermodynamics of crystals. Am. J. Phys. 40, 1718–1719.

Weerasinghe, G.L., et al., 2015. Computational searches for iron oxides at high pressure. J. Phys.: Condens. Matter 27, 455501.

Wentzcovitch, R.M., Martins, J.L., Allen, P.B., 1992. Energy versus free-energy conservation in first-principles molecular dynamics. Phys. Rev. B 45, 11372.

Wu, S.Q., Umemoto, K., Wentzcovitch, R.M., et al., 2011. Identification of post-pyrite

phase transitions in $SiO_2$ by a genetic algorithm. Physical Review B 83, 184102.

Wu, S.Q., et al., 2014. An adaptive genetic algorithm for crystal structure prediction. J. Phys.: Condens. Matter 26, 035402.

Zheng, F., et al., 2022. Structure and motifs of iron oxides from 1 to 3 TPa. Phys. Rev. Materials 6, 043602.

# Supplemental Material for “Novel phases in the Fe-Si-O system at terapascal pressures”

Nan Huang[1], Renata M. Wentzcovitch[2,3,4,5,†], Zepeng Wu[1], Feng Zheng[6], Bingxin Wu[1], Yang Sun[1,‡], Shunqing Wu[1,*]

*[1]Department of Physics, OSED, Key Laboratory of Low Dimensional Condensed Matter Physics (Department of Education of Fujian Province), Xiamen University, Xiamen 361005, China*

*[2]Department of Applied Physics and Applied Mathematics, Columbia University, New York, NY 10027, USA*

*[3]Department of Earth and Environmental Sciences, Columbia University, New York, NY 10027, USA*

*[4]Lamont–Doherty Earth Observatory, Columbia University, Palisades, NY 10964, USA*

*[5]Data Science Institute, Columbia University, New York, NY 10027, USA*

*[6]School of Science, Jimei University, Xiamen, 361021, China*

**Supplemental methods**

The structural prediction started with a general range search. In our methodology, we systematically explored the compositional space of the Fe-Si-O system by selecting distinct components characterized by the formula $Fe_xSi_yO_z$. The parameters x, y, and z are integers constrained to the set {1, 2, 3, 4} with their sums less than 10. Each combination was reduced to its simplest form, giving 46 unique permutations, which were used from 1 to a maximum of 10 formula units (not more than 30 atoms per cell) for the search. Considering that the reported high-pressure Mg-Si-O ternary compounds by Oganov (Niu et al., 2016) are all pseudo-binary in nature, we have also commenced a systematic exploration for prospective pseudo-binaries of the Fe-Si-O system. Selecting multiple pairs of edge points on the convex hull, we have conducted a search by combining the chosen endmembers' compositions in the ratios of 1:1, 1:2, 1:3, 2:1, and 3:1, respectively. The initial search did not yield stable structures; however, the results indicated that the formation enthalpies of the metastable structures in the oxygen-rich regions were generally lower. Following the initial exploration, we manually selected 20 compositions within the oxygen-rich region, maintaining the same formula unit configuration as in the first round. This search successfully identified a stable ternary phase of *P-3* $FeSi_2O_6$ and a metastable phase of *P1* $FeSi_4O_{12}$ closely approaching the convex hull with a formation enthalpy of 72 meV/atom. Both phases exhibit a high degree of local structural similarity.

To capitalize on the low-energy states afforded by their local structure, we conducted substitutions between iron and silicon. Following a comprehensive substitutional analysis and the elimination of duplicates, we proceeded with structural optimization. Interestingly, we achieved structures with superior energetic favorability. For *P-3* $FeSi_2O_6$, substitutional modifications yielded two novel ground states: *P3* $FeSiO_4$ and *P3* $Fe_4Si_5O_{18}$. However, under the newly established convex hull, *P3* $FeSi_2O_6$ no longer retains stability. In the case of *P1* $FeSi_4O_{12}$, a similarly energetically advantageous structure was identified as *P1* $Fe_3Si_7O_{24}$, exhibiting a $H_d$ of 69 meV/atom. This newly optimized structure lies closer to the convex hull by 2 meV/atom relative to *P1* $FeSi_4O_{12}$, yet it is still metastable.

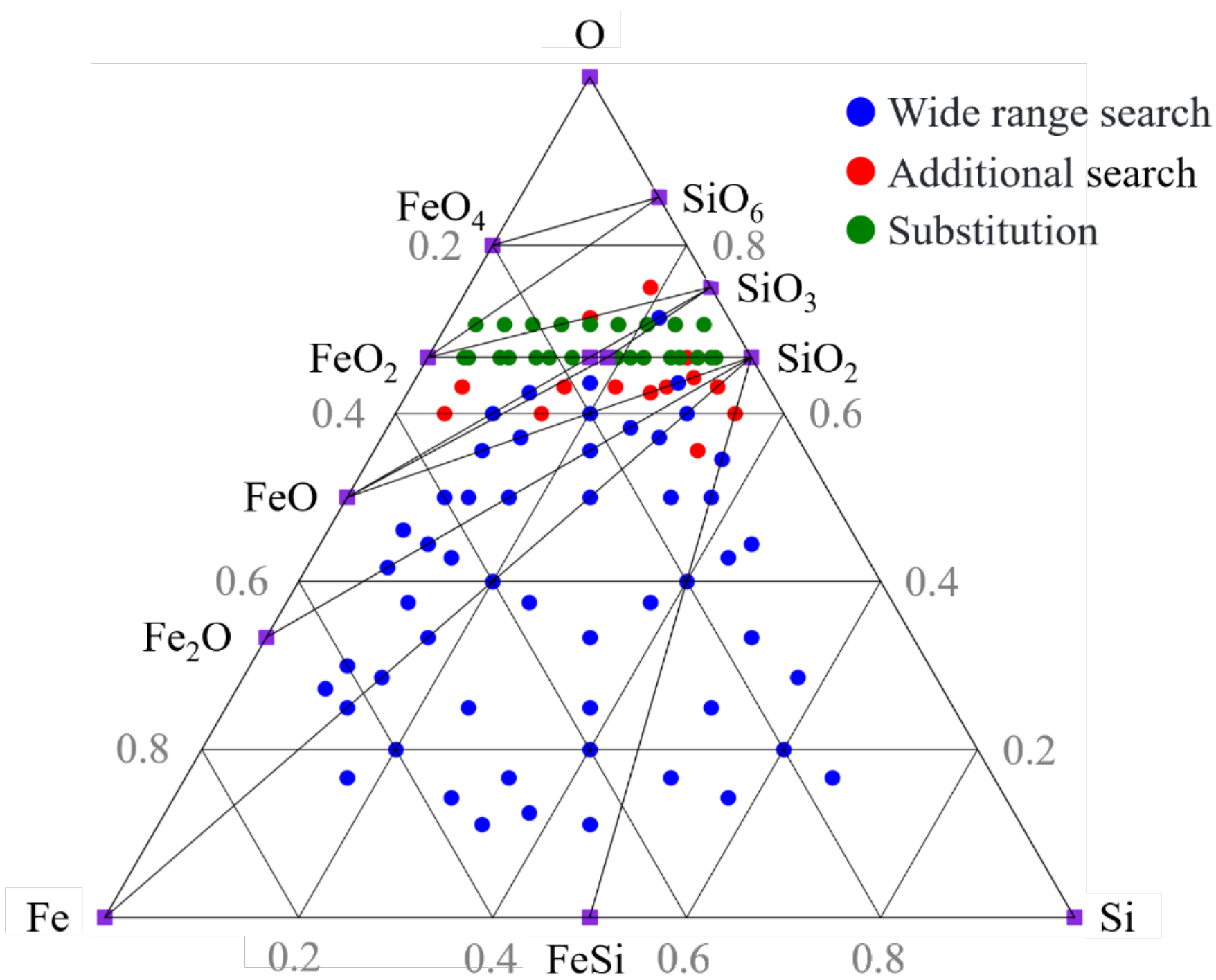

**Figure S1.** The classification of searched components. Blue, red, and green points represent a wide range, additional, and substitution search, respectively.

**Comparison of Pseudo-binary Convex Hulls**

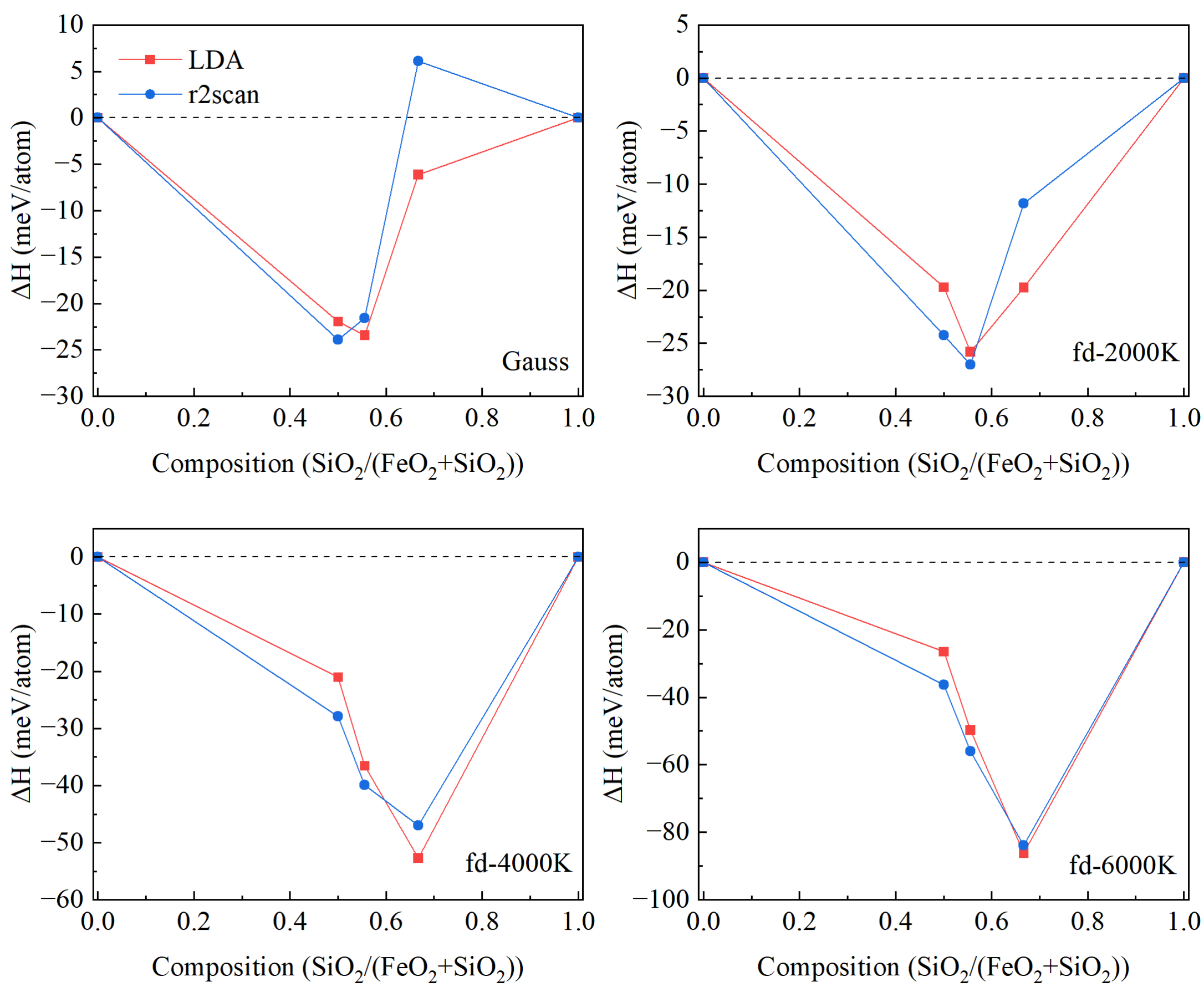

**Figure S2.** Pseudo-binary convex hulls were compared between QE (USPP-LDA) and VASP (PAW-r2SCAN), employing both Gaussian smearing and Fermi-Dirac smearing at various temperatures.

**Table S1.** Calculated space groups, lattice parameters, and Wyckoff positions of the investigated compounds in the Fe-Si-O system.

| Phase | SG | Lattice parameters | Wyckoff position |
|---|---|---|---|
| $FeSiO_4$(Z=3) | *P3* | a=4.219Å<br>b=4.219Å<br>c=4.103Å | Fe1 1b (0.333, 0.667, 0.509)<br>Fe2 1b (0.333, 0.667, 0.000)<br>Fe3 1c (0.667, 0.333, 0.621)<br>Si1 1a (0.000, 0.000, 0.880)<br>Si2 1a (0.000, 0.000, 0.375)<br>Si3 1c (0.667, 0.333, 0.123)<br>O1 3d (0.031, 0.751, 0.128)<br>O2 3d (0.034, 0.755, 0.628)<br>O3 3d (0.960, 0.370, 0.364)<br>O4 3d (0.954, 0.367, 0.876) |

| Phase | SG | Lattice parameters | Wyckoff position |
|---|---|---|---|
| $Fe_4Si_5O_{18}$(Z=1) | *P*3 | a=4.209Å<br>b=4.209Å<br>c=6.156Å | Fe1 1a (0.000, 0.000, 0.004)<br>Fe2 1a (0.000, 0.000, 0.335)<br>Fe3 1a (0.000, 0.000, 0.667)<br>Fe4 1c (0.667, 0.333, 0.081)<br>Si1 1b (0.667, 0.333, 0.254)<br>Si2 1b (0.667, 0.333, 0.585)<br>Si3 1b (0.667, 0.333, 0.916)<br>Si4 1c (0.667, 0.333, 0.416)<br>Si5 1c (0.667, 0.333, 0.747)<br>O1 3d (0.088, 0.700, 0.086)<br>O2 3d (0.085, 0.699, 0.420)<br>O3 3d (0.086, 0.699, 0.752)<br>O4 3d (0.922, 0.301, 0.251)<br>O5 3d (0.914, 0.299, 0.581)<br>O6 3d (0.923, 0.298, 0.909) |

| Phase | SG | Lattice parameters | Wyckoff position |
|---|---|---|---|
| $FeSi_2O_6$(Z=1) | $P\bar{3}$ | a=4.210Å<br>b=4.210Å<br>c=2.042Å | Fe1 1a (0.000, 0.000, 0.000)<br>Si2 2d (0.333, 0.667, 0.755)<br>O3 6g (0.086, 0.701, 0.257) |

## The phonon spectra, band-structure and projected DOS

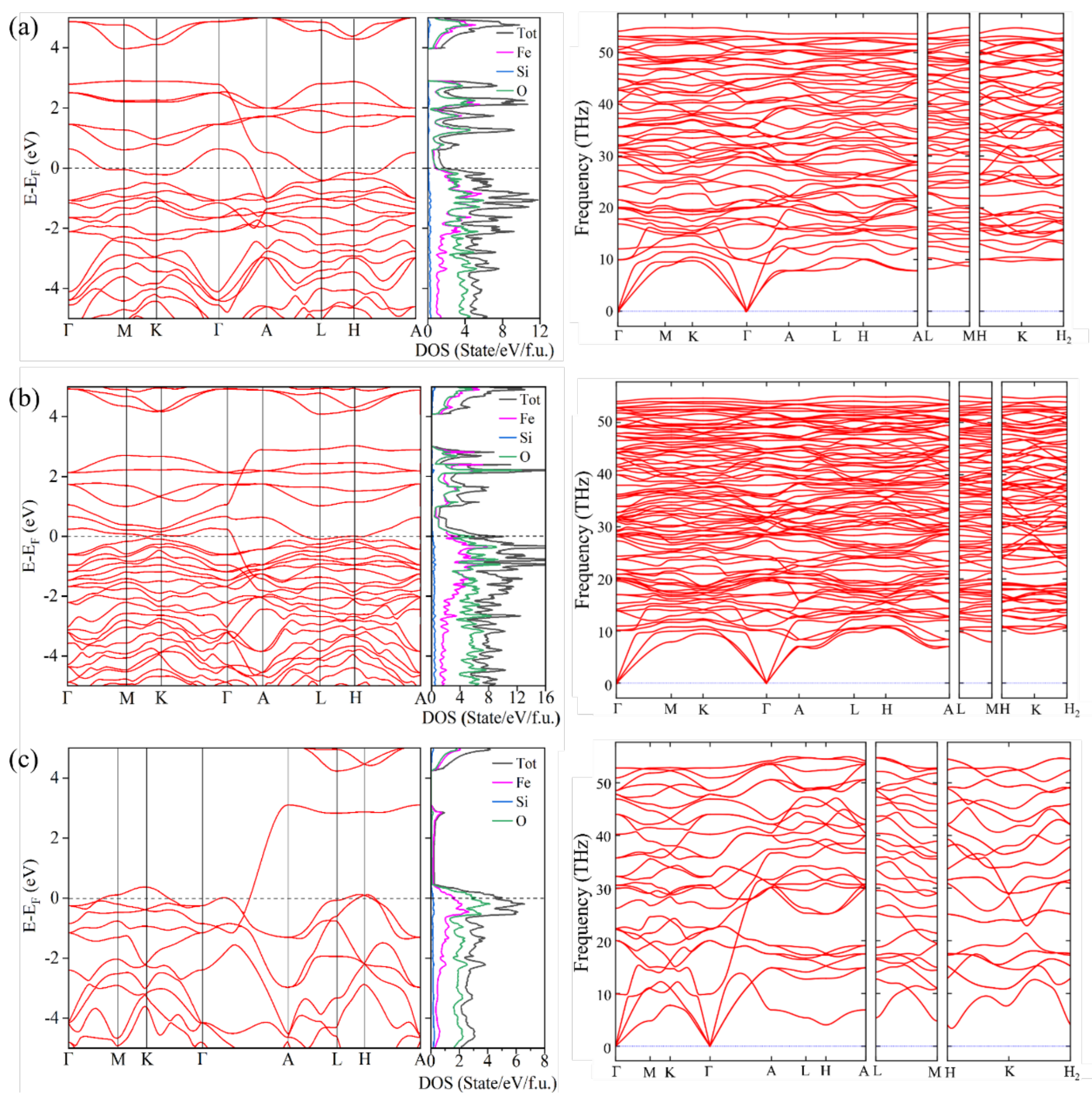

**Figure S3.** The phonon spectra, band-structure, and projected DOS of (a) *P3* $FeSiO_4$, (b) *P3* $Fe_4Si_5O_{18}$, and (c) $P\overline{3}$ $FeSi_2O_6$ at 1 TPa and 0 K. For projected DOS, the black, pink, blue, and green solid lines represent the total density of states and the partial wave densities of Fe, Si, and O, respectively.

## The complete structural evolution pathways to $P\bar{3}$ $FeSi_2O_6$

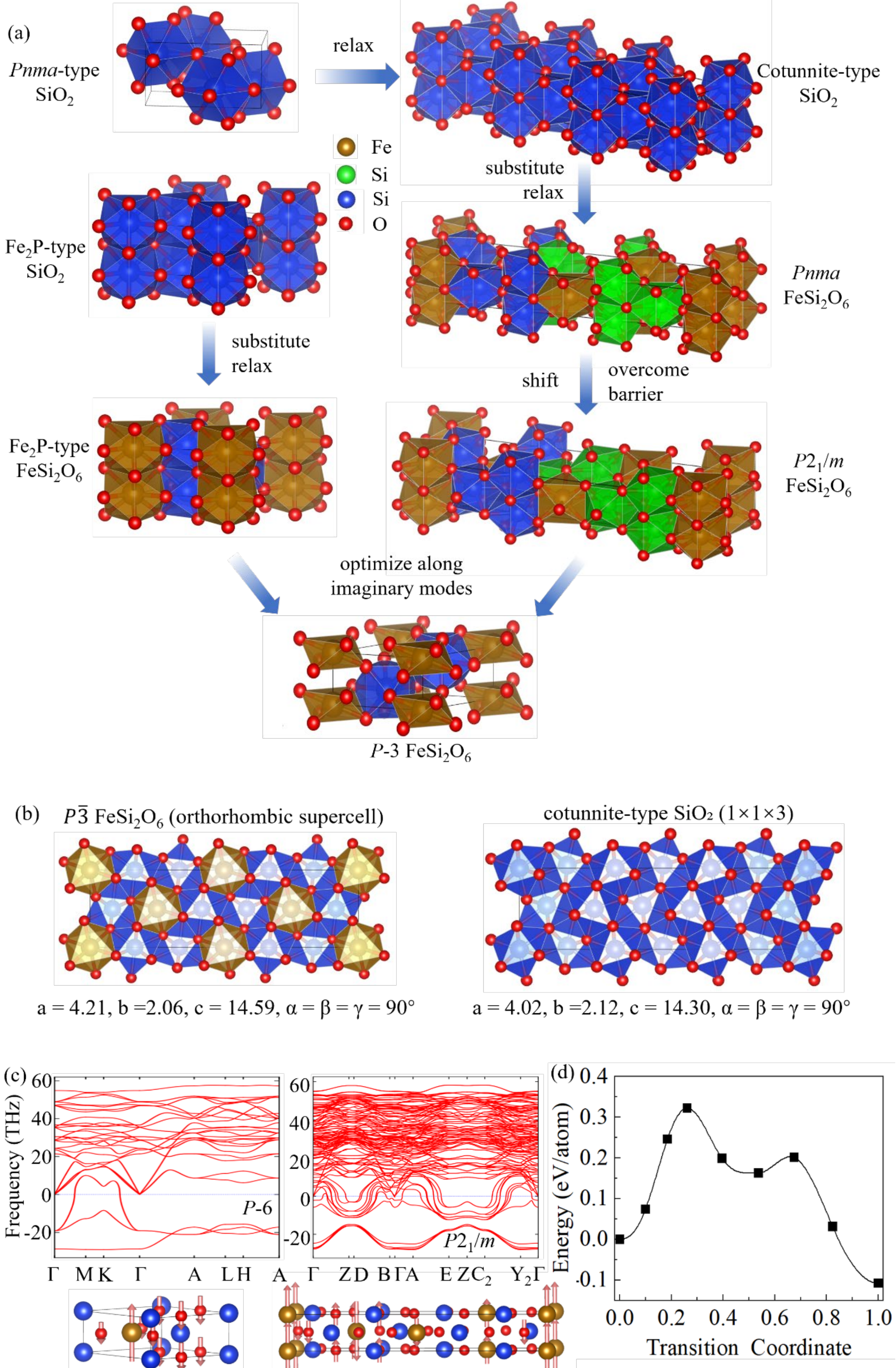

**Figure S4.** (a) The complete structural evolution pathways from $Fe_2P$-type $FeO_2$ and *Pnma*-type $SiO_2$ at 1 TPa to $P\overline{3}$ $FeSi_2O_6$. (b) Four-fold orthorhombic $FeSi_2O_6$ supercell and three-fold cotunnite-type $SiO_2$ supercell. (c) The phonon dispersion of $Fe_2P$-type (*P*-6) $FeSi_2O_6$ and $P2_1$/m $FeSi_2O_6$. The arrows on the atoms denote the directions of the selected unstable mode at the Γ point. (d) NEB calculation of the energy barrier for the *Pnma* to $P2_1/m$ transition in $FeSi_2O_6$. The green Si atoms shown in (a) are required to be displaced along the *b*-axis.

**Phase diagrams of FeO, $FeO_2$ and $SiO_2$**

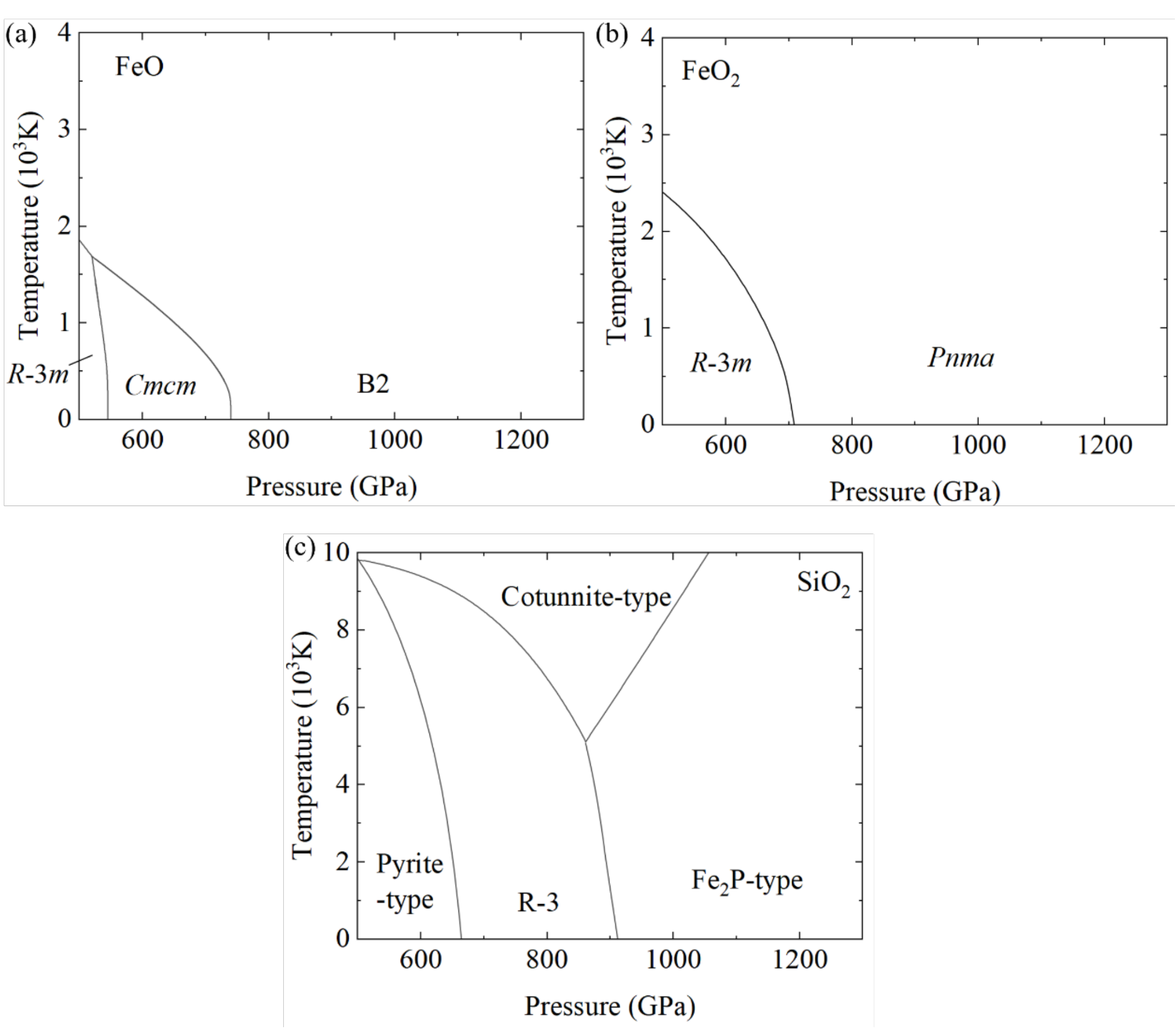

**Figure S5.** Phase diagrams of (a) FeO, (b) $FeO_2$, and (c) $SiO_2$. The phase boundaries are computed using Gibbs free energies computed using the QHA.

## References

Niu, H., Oganov, A.R., Chen, X.Q., Li, D., 2016. Prediction of novel stable compounds in the Mg–Si–O system under exoplanet pressures. Sci Rep 5, 18347.